\documentclass[superscriptaddress,prd,aps,preprint,showpacs]{revtex4}
\usepackage{graphicx,amsfonts,amsmath}

\begin{document}

\title{On the electrodynamics in time-dependent linear media}

\author{I. A. Pedrosa}
\email{iapedrosa@fisica.ufpb.br}
\author{A. Yu. Petrov}
\email{petrov@fisica.ufpb.br}
\author{Alexandre Rosas}
\email{arosas@fisica.ufpb.br}
\affiliation{Departamento de F\'{\i}sica, Universidade Federal da Para\'{\i}ba\\
 Caixa Postal 5008, 58051-970, Jo\~ao Pessoa, Para\'{\i}ba, Brazil}

\begin{abstract}
In this work we study the classical electrodynamics in homogeneous conducting and nonconducting time-dependent linear media in the absence of charge sources. Surprisingly, we find that the time dependence of the permittivity gives rise to an additional term in the Ampere-Maxwell equation and an asymmetry between the electric and magnetic field wave equations. As special cases we consider a linear and an exponential growth of the permittivity, as well as a sinusoidal time-dependent permittivity. 
\end{abstract}


\maketitle

\section{Introduction}

The fundamental problem of the behaviour of the electromagnetic field in material media has attracted a lot of attention of physicists for decades and continues to be a living and interesting research area of classical and quantum electrodynamics. The solution of this problem has supported the development of our understanding of the interaction of radiation and matter. The behaviour of the electromagnetic field is well understood in empty cavities or in free space but in the presence of the material media there is still work to be done.

In the past few years, the problem of electromagnetic waves propagating through dispersive and nondispersive material media has drawn a special attention of physicists motivated by the growth of experiments taking place inside material media \cite{1,2,3,4,5,6,7,8,9,10,11,12}. On the other hand, the study of electrodynamics in conducting and nonconducting time-dependent linear media (where the electric permittivity, magnetic permeability and/or conductivity vary in time) has also been an important subject of great physical interest \cite{13,14,15,16,17,18,19,20,21,22,23,24}. Here, it is worth mentioning that a time-dependent dielectric permittivity system can produce quanta of the electromagnetic field (photons) even from vacuum states \cite{13,14,15,16}. This phenomenon is similar to pure quantum effects such as dynamical Casimir effect (attractive interaction between two perfectly conducting plates separated by a short distance in vacuum) \cite{14,16}. Also, a model for plane wave in a medium with refraction index decreasing in time to study the Unruh effect (observers in an accelerating reference frame see a thermal radiation field at a temperature proportional to their acceleration) has been considered \cite{14,15}. Hence, the study of electrodynamics in nonstationary media is highly important both because of the practical applications and the academic interest.

In this work, we investigate the classical electromagnetic field in a homogeneous linear media, in the absence of charges, with time-dependent electric permittivity, magnetic permeability and conductivity. Remarkably, we find that the time dependence of the electric permittivity gives rise to an additional term in the right-hand side of the Ampere-Maxwell equation. Consequently we have obtained a generalization of the Ampere-Maxwell equation for nonstationary linear media. Furthermore, in this case, contrary to the time-independent one, the electric and magnetic fields do not satisfy the same wave equation, {\it i.e.}, the time-variation of the properties of the material media generates an asymmetry in the fields propagation. As particular illustrative cases, we consider linearly and exponentially growing electric permittivities, as well as a sinusoidally periodic one. 

We organize this paper as follows. In Section II, we investigate the electrodynamics in a homogeneous time-dependent linear media. In Section III we discuss the particular cases of time dependence of the electric permittivity mentioned above. We conclude the paper in Section IV with a brief summary.

\section{Electrodynamics in time-dependent linear media}  

We start our discussion with the well-known Maxwell's equations in media without charge sources, which are given by
\begin{eqnarray}
&&\vec{\nabla}\cdot\vec{D}=0, \label{maxwell1}\\
&&\vec{\nabla}\cdot\vec{B}=0, \label{maxwell2}\\
&&\vec{\nabla}\times\vec{E}=-\frac{\partial\vec{B}}{\partial t}, \label{maxwell3}\\
&&\vec{\nabla}\times\vec{H}=\vec{J}+\frac{\partial\vec{D}}{\partial t}. \label{maxwell4}
\end{eqnarray}
For time-dependent homogeneous linear media, the constitutive equations relating the fields and currents are given by
\begin{equation}
\vec{D}=\epsilon(t)\vec{E}, \quad
\vec{B}=\mu(t) \vec{H}, \quad
\vec{J}=\sigma(t) \vec{E}, \label{constitutive}
\end{equation}
where, $\epsilon(t), \;\mu(t)$ and $\sigma(t)$ are the time-dependent electric permittivity, magnetic permeability and conductivity, respectively. Then, using Eqs. (\ref{constitutive})  and (\ref{maxwell4})  we get
\begin{eqnarray}
\vec{\nabla}\times\vec{B}=\mu\vec{J}+\mu\epsilon\frac{\partial\vec{E}}{\partial t}+\mu\vec{E}\frac{\partial\epsilon}{\partial t}.
\end{eqnarray} 
This represents a generalization of the Ampere-Maxwell equation for nonstationary media with time-dependent electrical permittivity. Note that in the right-hand-side of this equation, the unusual rightmost term arises as a consequence of the time dependence of the electric permittivity. What is more, the time dependence of the magnetic permeability and of the conductivity do not generate any new term in the Ampere-Maxwell equation. As far as we know, the generalization of the Ampere-Maxwell equation for time-dependent media has not been considered in the literature yet. 

In order to write the wave equations for the electric and magnetic fields, we proceed as usual by taking the rotational of the Maxwell rotational equations [Eqs. (\ref{maxwell3}) and (\ref{maxwell4})]
\begin{eqnarray}
&&\vec{\nabla}\times(\vec{\nabla}\times\vec{E})=-\frac{\partial(\vec{\nabla}\times \vec{B})}{\partial t},\\
&&\vec{\nabla}\times(\vec{\nabla}\times\vec{B})=\mu \left [ \sigma\vec{\nabla}\times\vec{E}+\frac{\partial(\epsilon\vec{\nabla}\times\vec{E})}{\partial t}\right ].
\end{eqnarray}
Using the well known relation 
$\vec{\nabla}\times(\vec{\nabla}\times\vec{F})=\vec{\nabla}(\vec{\nabla}\cdot\vec{F})-\vec{\nabla}^2\vec{F},$ for a vector field $F$,
and remembering that in the absence of free charges $(\vec{\nabla}\cdot\vec{E})=(\vec{\nabla}\cdot\vec{B})=0$, we have
\begin{eqnarray}
&&\vec{\nabla}^2\vec{E}=\frac{\partial}{\partial t}\vec{\nabla}\times \vec{B}, \\
&&\vec{\nabla}^2\vec{B}=-\mu \sigma\vec{\nabla}\times\vec{E}-\mu\frac{\partial}{\partial t}\left ( \epsilon\vec{\nabla}\times \vec{E}\right ).
\end{eqnarray}
Now, taking into account the Eqs. (\ref{maxwell3}) and (\ref{maxwell4}) we obtain that
\begin{eqnarray}
\vec{\nabla}^2\vec{E}&=&\epsilon\mu\frac{\partial^2\vec{E}}{\partial t^2}+\mu\left (2\dot{\epsilon}\frac{\partial\vec{E}}{\partial t}+\ddot{\epsilon}\vec{E}\right )+\nonumber\\&+&\left (\dot{\epsilon}\dot{\mu}\vec{E}+\epsilon\dot{\mu}\frac{\partial\vec{E}}{\partial t}\right )+\sigma
\left (\dot{\mu}\vec{E}+\mu\frac{\partial\vec{E}}{\partial t}\right ) + \dot{ \sigma } \mu \vec{E},
\label{waveE1} \\
\vec{\nabla}^2\vec{B}&=&\epsilon\mu\frac{\partial^2\vec{B}}{\partial t^2}+\mu\left (\sigma+\dot{\epsilon}\right )\frac{\partial\vec{B}}{\partial t},\label{waveB}
\end{eqnarray}
where dots represent time derivatives. At this point it is already clear the asymmetry introduced by the time dependence of the electric permittivity, the magnetic permeability and the conductivity in the wave equations for electric and magnetic fields, that is, contrary to time-independent permittivity case, they do not satisfy the same wave equation. This notable property, to the best of our knowledge, has not been discussed in the literature yet. Furthermore, since the magnetic permeability usually almost does not differ from the vacuum value $\mu_0$, except for ferromagnets, we make the approximations $\dot{\mu}\simeq 0$ and $\ddot{\mu}\simeq 0$. This approximation is valid even for ferromagnets provided that the magnetic permittivity is constant. As a result, we have
\begin{equation}
\vec{\nabla}^2\vec{E}=\epsilon\mu\frac{\partial^2\vec{E}}{\partial t^2}+\mu(2\dot{\epsilon} + \sigma )\frac{\partial\vec{E}}{\partial t}+\mu(\ddot{\epsilon} +\dot{ \sigma })\vec{E}.\label{waveE}
\end{equation}
It is worth noticing that the equivalent equation for the magnetic field is not modified under these approximations. 
What is more, from Eqs. (\ref{waveE}) and (\ref{waveB}) we see that these equations differ by a factor 2 which multiply the first time-derivative of the permittivity and the additional terms proportional to the electric field, which appear only in the electric field wave equation. Thus, the asymmetry between electric and magnetic fields remains even if the magnetic permeability is time-independent.
Further, since the vector potential and the magnetic field relation is time-independent, the wave equation for the vector potential in the Coulomb gauge is exactly the same as the one for the magnetic field~\cite{21}.

Now, let us look for solutions to the wave equations for $\vec{E}$ and $\vec{B}$. Employing the method of separation of variables, we write the electric field as \cite{17,21,25}
\begin{eqnarray}
\label{repr}
\vec{E}(\vec{r},t)=\sum\limits_{l}\vec{u}_l(\vec{r})q_l(t),
\end{eqnarray}
where $\vec{u}_l(\vec{r})$ and $q_l(t)$ are mode and amplitude functions (we consider the electromagnetic field in a certain volume of space). Substituting (\ref{repr}) into (\ref{waveE}) we find that
\begin{equation}
\sum\limits_{l}\left[\vec{\nabla}^2\vec{u}_l(\vec{r})q_l(t)-\mu(2\dot{\epsilon}+\sigma)\vec{u}_l(\vec{r})\dot{q}_l(t)-\mu\epsilon\vec{u}_l(\vec{r})\ddot{q}_l(t) - \mu ( \ddot{ \epsilon } + \dot{ \sigma })\vec{u}_l(\vec{r})q_l(t)
\right] =0.
\end{equation}
Thus, for any component $i$ of any vector $\vec{u}_l$ we have
\begin{equation}
\label{separation}
\frac{\vec{\nabla}^2u_{il}(\vec{r})}{u_{il}(\vec{r})}=\frac{\mu (\ddot{ \epsilon } + \dot{ \sigma }) q_l(t) +\mu(2\dot{\epsilon}+\sigma)\dot{q}_l(t)+\mu\epsilon\ddot{q}_l(t)
}{q_l(t)}.
\end{equation}
The left and right hand sides of Eq. (\ref{separation}) must be equal to the same constant, since they are function of the position and time, respectively. Therefore, we have,
\begin{eqnarray}
\label{u1}
&&\vec{\nabla}^2\vec{u}_l(\vec{r})+\frac{\omega^2_l}{c^2_0}\vec{u}_l(\vec{r})=0,\\
\label{u2}
&&\ddot{q}_l+\frac{2\dot{\epsilon}+\sigma}{\epsilon}\dot{q}_l+\left ( \frac{\ddot{ \epsilon } + \dot{ \sigma }}{ \epsilon }+\Omega_l^2(t) \right ) q_l=0,
\end{eqnarray}
where
\begin{equation}
\Omega^2_l(t)=\frac{c^2(t)\omega^2_l}{c^2_0}.
\end{equation}
Here $c(t)=\left[\mu\epsilon(t)\right]^{-1/2}$ is the speed of light in the medium, $(\omega_l^2/c^2_0)$ is the eigenvalue of the Laplacian of $u_l(\vec{r})$ and $c_0=\left[\mu_0\epsilon_0\right]^{-1/2}$ is a speed of light in the vacuum.

Now, let us proceed with similar methodology for the magnetic field and write
\begin{equation}
\label{reprt}
\vec{B}(\vec{r},t)=\sum\limits_{l}\vec{u}_l(\vec{r})\tilde{q}_l(t),
\end{equation}
where $\tilde{q}_l(t)$ are the amplitude functions (once again, we consider the electromagnetic field in a certain volume of space). Proceeding along the same lines as above, we find the following equation for the amplitude function
\begin{equation}
\label{u3}
\ddot{\tilde{q}}_l+\frac{\dot{\epsilon}+\sigma}{\epsilon}\dot{\tilde{q}}_l+\Omega_l^2(t)\tilde{q}_l=0.
\end{equation}
From the equations (\ref{u2}) and (\ref{u3}) we see that for $\dot{\epsilon}>0$ the amplitudes of the electric and magnetic fields are damped. Thus, the time dependence of the electric permittivity also causes an additional damping in the amplitudes of the electric and magnetic fields. This interesting effect has already been considered in Ref. \cite{21}.

In the following, we move our attention to the solutions of (18). Considering the  electromagnetic field to be contained in a certain cubic volume $V$ of nonrefracting media, the mode functions are required to satisfy the transversality condition $\vec{\nabla}\cdot\vec{u_l}=0$ and to form a orthonormal set \cite{25}. Furthermore, assuming periodic boundary conditions on the surface, the mode function may be written in terms of plane waves as \cite{17,21,25}
\begin{equation}
\vec{u}_{l\nu}(\vec{r}) = L^{-3/2} e^{\pm i \vec{k}_l\cdot \vec{r}} \hat{e}_{l\nu},
\end{equation}
where $L=V^{1/3}$ is the size of the cube, $|\vec{k_l}|= \omega_l/c_0$ is the wave vector and $\hat{e}_{l\nu}$ are unit vectors in the directions of polarization ($\nu=1,2$), which must be perpendicular to the wave vector because of the transversality condition. With the spacial mode functions $\vec{u_l}$ completely determined, we only need the canonical variable $q_l(t)$ in order to obtain the electric field $\vec{E}$. Then, the field $\vec{E}$ confined in the cubic volume of side $L$ can be written as
\begin{equation}
\vec{E}(\vec{r},t)= \frac{1}{ L^{3/2}}\sum_l\sum_{\nu}\hat{e}_{l\nu}e^{\pm i\vec{k}_l\cdot\vec{r}}q_l(t).
\end{equation}

Analogously, the magnetic field $\vec{B}$ takes the form
\begin{equation}
\vec{B}(\vec{r},t)= \frac{1}{ L^{3/2}}\sum_l\sum_{\nu}\hat{e}_{l\nu}e^{\pm i\vec{k}_l\cdot\vec{r}}\tilde{q_l}(t).
\end{equation}
Thus, with the knowledge of $q_l(t)$ and $ \tilde{q}_l(t)$, which are obtained from Eqs. (\ref{u2}) and (\ref{u3}), respectively, we have a complete description of the electromagnetic field.

\section{Time-varying permittivity: special cases}

\begin{figure}
\begin{center}
  \includegraphics[width = 10cm]{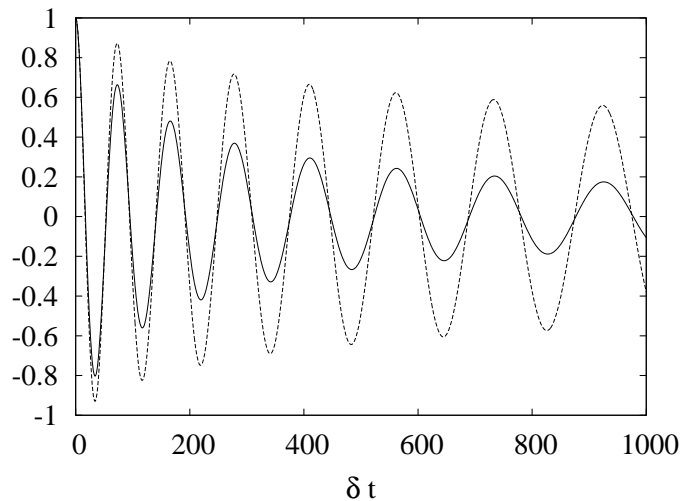}
\caption{Time evolution of the amplitude function of a mode of the electric  (full line) and magnetic (dashed line) fields for a linear growth of the electric permittivity. The initial condition was set as $q_l(0) = \tilde{q}(0) = 1$ and $\dot{q}_l(0) = \tilde{\dot{q}}(0) = 0.$ The ratio of the parameters $ \omega_l$ and $ \delta$ was chosen to be 10. \label{fig:linear}}
\end{center}
\end{figure}

\begin{figure}
\begin{center}
  \includegraphics[width = 10cm]{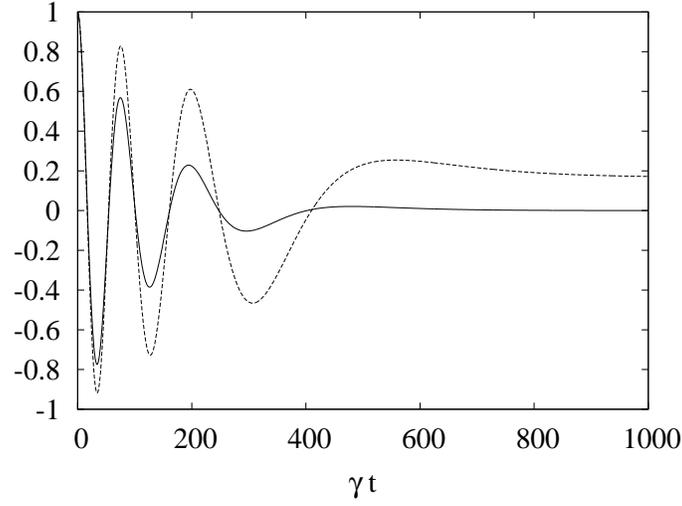}
\caption{Time evolution of the amplitude function of a mode of the electric  (full line) and magnetic (dashed line) fields for a exponential growth of the electric permittivity. The initial condition was set as $q_l(0) = \tilde{q}(0) = 1$ and $\dot{q}_l(0) = \tilde{\dot{q}}(0) = 0.$ The ratio of the parameters $ \omega_l$ and $ \gamma$ was chosen to be 10. \label{fig:exp}}
\end{center}
\end{figure}

In this section, we discuss the time-evolution of the electric and magnetic fields for different time dependences of the electric permittivity. For simplicity, we restrict our analysis to the case of dielectric materials ($ \sigma = 0$) and non-ferromagnetic materials ($ \mu \simeq \mu_0$). The first case we consider is the one of an electric permittivity which grows linearly with time, i.e. $\epsilon(t)=\epsilon_0 (1+\delta t),$ where $ \delta $ is a constant with units of the inverse of time. In this case, the amplitude functions for the electric and magnetic fields, Eqs. (\ref{u2}) and (\ref{u3}), respectively, become
\begin{eqnarray}
&&\ddot{q}_l+\frac{ 2\delta }{ 1 + \delta t}\dot{q}_l+\frac{\omega_l^2}{1+\delta t}q_l=0.\\
&&\ddot{\tilde{q}}_l+\frac{ \delta }{ 1 + \delta t}\dot{\tilde{q}}_l+\frac{\omega_l^2}{1+\delta t}\tilde{q}_l=0.
\end{eqnarray}
In both cases, the solution is given in terms of the $I_k$ and $K_k$ Bessel functions as \cite{26}
\begin{eqnarray}
q_l(t) &=&
\frac{ c_1 \omega_l  }{\delta  \sqrt{\delta  t+1}}I_1\left(\frac{2 i \sqrt{t \delta +1} \omega_l }{\delta
   }\right)
-\frac{ c_2 \omega_l  }{\delta  \sqrt{\delta  t+1}}K_1\left(\frac{2 i \sqrt{t \delta +1} \omega_l }{\delta
   }\right)\\
\tilde{q}_l(t) &=&
\frac{c_3 }{\sqrt{\delta }} I_0\left(\frac{2 i \sqrt{t \delta +1} \omega_l }{\delta
   }\right)+
\frac{c_4 }{\sqrt{\delta }}K_0\left(\frac{2 i \sqrt{t \delta +1} \omega_l }{\delta
   }\right)
\end{eqnarray}
where $c_{k}$ are constants to be determined by the initial conditions. As expected, both field vanish with $t$ due to the electric permittivity increase. However, they decay with distinct rates. While the electric field decays, in the long time regime, as $t^{-3/4}$, the magnetic field decay is slower ($t^{-1/4}$). It is worth noticing that, for a linear growth of the electric permittivity, the only difference between the electric and magnetic amplitude function equation is the factor $2$ in the first derivative of the amplitude function. In Fig.~\ref{fig:linear}, we show a typical time evolution of the electric and magnetic field, where the faster decay of the electric field is clear. This figure also illustrate the oscillatory behavior of both fields.
Now, let us consider a exponentially growing electric permittivity $ \epsilon (t) = \epsilon _0 e^{\gamma t},$ where $ \gamma $ is a constant with units of the inverse of time. In this case, the amplitude functions for the electric and magnetic fields are respectively
\begin{eqnarray}
&& \ddot{q}_l+2 \gamma  \dot{q}_l +q_l e^{-\gamma t} \left(\gamma ^2 e^{\gamma t}+\omega_l ^2\right) = 0.\\ 
&& \tilde{\ddot{q}}_l+\gamma  \tilde{\dot{q}}_l+\omega_l ^2 \tilde{q}_l e^{-\gamma t} = 0.
\end{eqnarray}
In this case, there is an additional term proportional to the amplitude function which comes from the second derivative of the electric permittivity. Once again, the solutions may be written in terms of $J_k$ and $Y_k$ Bessel functions~\cite{26}
\begin{eqnarray}
q_l(t) &=& \frac{\omega_l ^2 e^{-\gamma t} }{\gamma ^2}\left[d_1 J_0\left(\frac{2 e^{-\frac{t
   \gamma }{2}} \omega_l }{\gamma }\right)-2 d_2 Y_0\left(\frac{2 e^{-\frac{t
   \gamma }{2}} \omega_l }{\gamma }\right)\right],\\ 
\tilde{q}_l(t) &=& \frac{\omega_l  e^{-\frac{\gamma  t}{2}} }{\gamma }\left[d_3 J_1\left(\frac{2
   e^{-\frac{t \gamma }{2}} \omega_l }{\gamma }\right)+2  d_4
   Y_1\left(\frac{2 e^{-\frac{t \gamma }{2}} \omega_l }{\gamma
   }\right)\right],
\end{eqnarray}
where $d_k$ are constants to be determined by the initial conditions. The asymptotic behavior for large $t$ is even more surprisingly. While the $J_0, \; Y_0$ and $J_1$ terms all decay to zero exponentially, as $e^{- \gamma t}$, the $Y_1$ term which appears in the solution of the amplitude functions of the magnetic field goes to $2 d_4/ \pi.$ Hence, depending on the initial conditions, the long time behavior may be a non-zero constant magnetic field, as can be seen in Fig.~\ref{fig:exp}. Comparing Figs.~\ref{fig:linear} and~\ref{fig:exp}, we can see that the decay of the electromagnetic field is faster in the later case.

\begin{figure}
\begin{center}
  \includegraphics[width = 10cm]{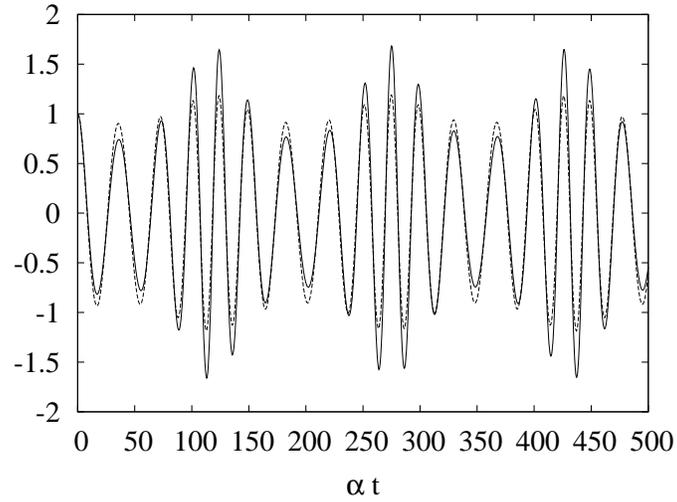}
\caption{Time evolution of the amplitude functions of a mode of the electric  (full line) and magnetic (dashed line) fields for a periodic electric permittivity. The initial condition was set as $q_l(0) = \tilde{q}(0) = 1$ and $\dot{q}_l(0) = \tilde{\dot{q}}(0) = 0.$ The ratio of the parameters $ \omega_l$ and $ \alpha$ was chosen to be 20. \label{fig:sin}}
\end{center}
\end{figure}

Next, let us assume a sinusoidal time-varying permittivity given by $ \varepsilon (t) = \varepsilon_0 \left [ 1 + B \sin ( \alpha t ) \right ],$ where $ \gamma $ is a constant with units of the inverse of time and $B$ in an dimensionless constant. In this case, we were unable to find a analytical solution. Hence, we solved the differential equations for the amplitude function numerically. As shown in Fig.~\ref{fig:sin}, the amplitude functions for the electric and magnetic fields oscillate in phase, but they still differ (while for time-independent electric permittivity, they are the same).
 
\section{Summary}

In this work, we have presented a simple and clean-cut treatment for the problem of classical electrodynamics in nonstationary media without charge sources. We have seen that the time dependence of the electric permittivity gives rise to an additional term in the right-hand-side of the Ampere-Maxwell equation. Besides, the time dependence of the electric permittivity, the magnetic permeability and the conductivity cause an asymmetry between the electric and magnetic field wave equations. These surprisingly and intriguing results have not been explored in the literature yet. We also have considered as particular cases a linearly and an exponentially increasing permittivity, as well as a sinusoidal periodic electric permittivity. We showed that in the first two case the electromagnetic field was attenuated, while the sinusoidal time dependence induced a periodic behavior of the fields. Furthermore, the quantum counterpart of this problem (for real electric permittivity and magnetic permeability) can, in principle, be  carried out by following step by step the procedure developed in Refs. \cite{21,23}. Finally, we would like to mention that further study on electrodynamics in nonstationary linear media is in progress and the results will be published elsewhere. Yet we expect that the results obtained in this work can be useful to other researchers investigate subjects related to the electrodynamics in conducting and nonconducting  media with material properties varying in time.

{\bf Acknowledgments.}
 The authors were partially supported by CNPq. I. A. Pedrosa and A. Rosas also thank to Capes/nanobiotec for partial financial support.

\end{document}